\documentclass[conference,letterpaper]{IEEEtran}
\IEEEoverridecommandlockouts
\usepackage{cite}
\usepackage{amsmath,amssymb,amsfonts,amsthm}
\usepackage{algorithmic}
\usepackage{graphicx}
\graphicspath{{fig/}}
\usepackage[caption=false,font=footnotesize]{subfig}
\usepackage{textcomp}
\usepackage{xcolor}
\usepackage{soul}
\input{Qcircuit}
\usepackage{braket}
\def\BibTeX{{\rm B\kern-.05em{\sc i\kern-.025em b}\kern-.08em
    T\kern-.1667em\lower.7ex\hbox{E}\kern-.125emX}}
\begin{document}

\title{Study of Decoherence in Quantum Computers: \\A Circuit-Design Perspective
}

\author{\IEEEauthorblockN{Abdullah Ash- Saki}
\IEEEauthorblockA{\textit{Electrical Engineering} \\
\textit{Pennsylvania State University}\\
University Park, USA \\
ash.saki@psu.edu}
\and
\IEEEauthorblockN{Mahabubul Alam}
\IEEEauthorblockA{\textit{Electrical Engineering} \\
\textit{Pennsylvania State University}\\
University Park, USA \\
mxa890@psu.edu}
\and
\IEEEauthorblockN{Swaroop Ghosh}
\IEEEauthorblockA{\textit{Electrical Engineering} \\
\textit{Pennsylvania State University}\\
University Park, USA \\
szg212@psu.edu}

\thanks{\textbf{Presented in GOMACTech 2019, March 25-28, 2019, Albuquerque, NM}}
}

\maketitle

\begin{abstract}
Decoherence of quantum states is a major hurdle towards scalable and reliable quantum computing. Lower decoherence (i.e., higher fidelity) can alleviate the error correction overhead and obviate the need for energy-intensive noise reduction techniques e.g., cryogenic cooling. In this paper, we performed a noise-induced decoherence analysis of single and multi-qubit quantum gates using physics-based simulations. The analysis indicates that (i) decoherence depends on the input state and the gate type. Larger number of $\ket{1}$ states worsen the decoherence; (ii) amplitude damping is more detrimental than phase damping; (iii) shorter depth implementation of a quantum function can achieve lower decoherence. Simulations indicate 20\% improvement in the fidelity of a quantum adder when realized using lower depth topology. The insights developed in this paper can be exploited by the circuit designer to choose the right gates and logic implementation to optimize the system-level fidelity.

\end{abstract}

\section{Introduction}
In 1982, Richard P. Feynman envisioned that to simulate nature we would need a quantum mechanical computer \cite{fynman}. Since then different industrial and academic groups have worked diligently to realize a physical quantum computer. The building block of the quantum computer is called a quantum bit or qubit. Different groups have come up with different physical realization of qubits including ion-trap \cite{iont}, superconducting circuits \cite{supercond}, semiconductor quantum dots \cite{qdots}, a single atom in Silicon \cite{silicon} etc. each with its own benefit and caveat. Qubit possesses some unique properties like superposition, entanglement and quantum interference. Due to these unique properties’ quantum computers are prophesied to efficiently solve some problems which are thought to be intractable for classical computers. Already several quantum algorithms are proposed including Shor's prime factorization, Grover's search, Harrow-Hassidim-Lloyd's linear system of equations which exhibit superior performance than their classical counter-part and thus, pushing towards the goal of quantum supremacy \cite{supremacy}. The number of qubits is increasing with a 128-qubit quantum computer is expected by 2019 \cite{riggeti}. More qubits generally mean more computational power. 

However, with existing technologies, qubit states are short lived. They tend to \textit{decohere} due to different noise sources and interaction with environment \cite{joos}. The decoherence problem worsens with an increasing number of qubits. To mitigate those, techniques like cryogenic cooling and error correction code \cite{errorcc} are employed. However, these techniques are costly as cooling the qubits to ultra-low temperature (mili-Kelvin) requires very high power (as much as 25kW) and error correction requires many redundant physical qubits (one popular error correction method named \textit{surface code} incurs 18X overhead \cite{surfacecode}). 

Circuit level implications of decoherence are not well understood. Therefore, a circuit level analysis of decoherence process is warranted to provide insights that may open avenues for more optimization and stabilization techniques and thus, enable the promise of quantum computers with the available noisy-intermediate-scale quantum (NISQ) \cite{nisq} technology. To this end, the following contributions are made in this paper: 
\begin{itemize}
\item We explore multi-qubit and multi-depth quantum circuits from a circuit-design standpoint. 
\item We employ a Physics based Master equation framework to model the dynamics of the qubit state with environmental interaction (noise).
\item We report a trend in decoherence with input pattern. 
\item We show that lower-depth quantum circuits are better performing in terms of decoherence than higher-depth circuits. 
\end{itemize}

The rest of the paper is organized as follows: in Section II, we discuss the basics of quantum computing and describe the simulation framework used in the paper. In Section III, we present result for single quantum gates. In Section IV, we extend the analysis to multi-gate circuits with a test case of two different implementations of a quantum full adder. Finally, we draw a conclusion in Section V. 

\section{Basics of Quantum Computing and Simulation Framework}
In this section, first, we introduce basic terminologies of quantum computing and then discuss the simulation framework and underlying assumptions used in this paper. 
\subsection{Basic Terminologies}\label{basics}
\begin{figure}[b]
\vspace{-5mm}
\centerline{\includegraphics[width=1.5in]{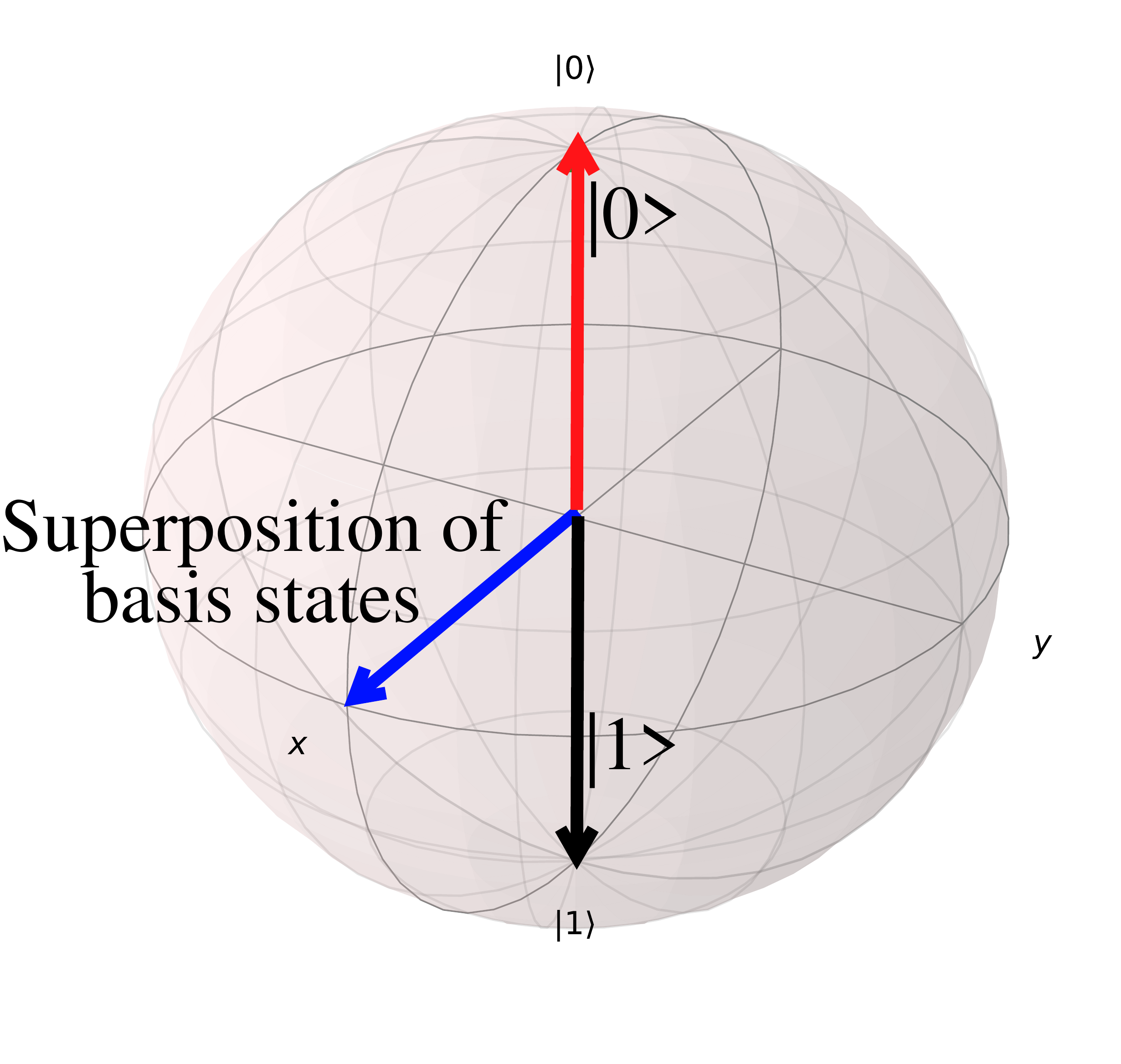}}
\caption{Bloch sphere representation of qubit state.}
\label{bloch}
\end{figure}

\textbf{State vector:} Quantum computer comprises of qubits which are the quantum analogue for classical bits. The state of a qubit at a time instance is given by the \emph{state vector} $\ket{\psi}$. Generally, $\ket{\psi}$ can be expressed as $\ket{\psi} = a \ket{0} + b \ket{1}$
%
where, $a$ and $b$ are complex numbers such that $|a|^2 + |b|^2 = 1$. $\ket{0}$ and $\ket{1}$ are orthonormal set of basis vectors which span the state space of a physical system (also known as \emph{Hilbert space} in the context of quantum mechanics). 
Qubit can take a state which is neither 0 nor 1, rather a \emph{superposition} of both states. $\ket{\psi} = (\ket{0} + \ket{1})/\sqrt{2}$ is an example of such state that has $50\%$ probability of being $\ket{0}$ and $50\%$ probability of being $\ket{1}$. Fig. \ref{bloch} graphically depicts the state vectors $\ket{0}$, $\ket{1}$ and their superposition state. 

\textbf{Density matrix:} Besides state vector, state of a qubit can be described with another entity known as \emph{density operator} or \emph{density matrix} which is defined by
\begin{equation}\label{dm}
\rho \equiv \sum_i p_i\ket{\psi_i}\bra{\psi_i} 
\vspace{-2mm}
\end{equation} 
where $\ket{\psi_i}$ is one of the numbers of states of the quantum system and $p_i$ is the probability of that state. 
Although this alternate description is mathematically equivalent to state vector formulation, it provides convenient means of describing the (dynamic) evolution of a quantum system, especially systems with noise.       

\textbf{Time evolution and Hamiltonian:} During quantum computation, qubit states evolve with time as governed by \emph{Schr\"{o}dinger equation}. The evolution is a \emph{unitary transformation}. If the state $\ket{\psi}$ of a quantum system is evolved from time $t_1$ to $t_2$, then the evolution is described by \emph{Schr\"{o}dinger equation} as follows:
\begin{align*}
i\hbar\frac{d\ket{\psi}}{dt} &= H\ket{\psi}\\
\therefore \ket{\psi(t_2)} &=exp(-i \frac{H}{\hbar} t) \ket{\psi(t_1)}\\
\text{or,}\hspace{0.2cm} \rho(t_2) &= U\rho(t_1)U^{\dagger}\\
[\text{where, }U &= exp(-i \frac{H}{\hbar} t), t = t_2 - t_1]
\end{align*}
$H$ is a \emph{Hermitian} operator known as the system \emph{Hamiltonian} which dictates the evolution of the closed system. $U$ is the corresponding unitary matrix for Hamiltonian $H$. In gate-based quantum computing regime, different gate operation is defined by different unitary matrices. For example, quantum NOT gate is defined by the following matrix ($U_{NOT}$) as it converts $\ket{\psi} = \ket{0}$ to $\ket{1}$ and vice versa. 
\begin{align*}
U_{NOT} = \begin{bmatrix}0 & 1\\1 & 0\\\end{bmatrix} \hspace{1cm} \text{and,} \hspace{1cm}\begin{bmatrix}1\\0\\\end{bmatrix} &= \begin{bmatrix}0 & 1\\1 & 0\\\end{bmatrix} \begin{bmatrix}0\\1\\\end{bmatrix}
\end{align*}
As Hamiltonian governs the evolution of qubit state, it is of paramount importance. At present, there are a number of different physical devices to realize quantum computers e.g. Ion trap \cite{iont}, superconducting qubit \cite{supercond}, Silicon-based \cite{silicon}, Nuclear Magnetic Resonance (NMR) based etc. quantum computers. Each physical realization has its own definition of the  Hamiltonian. For example, the Hamiltonian for a closed $n$ spin coupled NMR system is given by \cite{chuang}:
\begin{equation}
H = \sum_k\omega_kZ_k + \sum_{j,k}H_{j,k}^J + \sum_{j,k}H_{j,k}^D + H^{RF}
\end{equation}
The first three terms describe the free precession of the spins in the ambient magnetic field, the magnetic dipole coupling of the spins and the $J$ coupling of the spins respectively (a more complete description of these terms can be found in \cite{chuang}). However, the $H^{RF}$ term is particularly interesting from an external control standpoint. In presence of a large radio-frequency (RF) field of a proper frequency, the unitary evolution of the system can be approximated as \cite{chuang}:
\begin{equation}
exp(-i\frac{H}{\hbar}t) \approx exp(-i\frac{H^{RF}}{\hbar}t) \label{Happrox}
\end{equation}
Equation \eqref{Happrox} provides the insight that using external fields one can control the evolution (unitary transformation) of the quantum system in an arbitrary way (although, different physical realization imposes different constraints to this statement \cite{chuang}. However, each realization has at least a set of universal gates (unitary transformations) that can be implemented through external control such as RF field in case of NMR quantum computers).  

\textbf{Open system and Master equation}: Although Schr\"{o}dinger equation can describe the evolution of closed quantum systems, it is not sufficient for \emph{open systems}. An open system is a quantum system that interacts with the environment. In reality, there are no perfectly isolated system and every system is basically an open system. Interaction with environment portrays as \emph{noise} in quantum computing and leads to the quantum phenomenon named \emph{decoherence}. An intuitive description of the problem related to decoherence can be given using \textit{Schr\"{o}dinger's cat} analogy. The qubit can be at a superposition of both $\ket{0}$ and $\ket{1}$ states simultaneously. However, the environment works as an observer and according to the laws of quantum physics, this observation forces one of the states to collapse leading to a classical state of either $\ket{0}$ or $\ket{1}$. This collapse of superposition state is a reason behind decoherence. 

The dynamics of an open quantum system can be described mainly with two methods. One approach is using the operator-sum method with Kraus operators \cite{kraus}.  An alternate approach is to use the master equation, which can be written most generally in the \emph{Lindblad form} \cite{lind76} as: 
\begin{equation}
\vspace{-2mm}
\frac{d\rho}{dt} = - \frac{i}{\hbar} [H, \rho] + \sum_{j} [2L_j \rho L_{j}^{\dagger} - L_{j}^{\dagger}L_j\rho - \rho L_{j}^{\dagger} L_j] \label{mastereq}
\end{equation}
where $H$ is the system Hamiltonian and,
$L_j$ are the \emph{Lindblad} operators which represent the coupling of the system to its environment. This equation considers the Born-Markov approximation \cite{bre}.
Solving \eqref{mastereq} will give the density matrix $\rho$ at different time instances. In this paper, we use Master equation as it can provide continuous time evolution. 

\textbf{Fidelity}: Due to decoherence qubits lose their states. Therefore, the qubit state after a gate operation in presence of noise (i.e. with decoherence), may not be equal to the target state in absence of noise. Therefore, a metric named \textit{fidelity} is widely used to quantify the reliability of quantum operation(s). 
For example, states $\ket{\psi}=0$ and $\ket{\psi}=1$ are completely opposite. Therefore, the fidelity between these states should be 0. In mathematical terms, fidelity between two states $\rho$ and $\sigma$ is defined by \cite{chuang}:
\begin{equation}
F(\rho, \sigma) \equiv tr(\sqrt{\rho^{1/2}\sigma\rho^{1/2}})
\vspace{-2mm}
\end{equation}
where $tr(.)$ is the matrix trace operation. 
\subsection{Simulation Setup}

We have used Qutip \cite{qutip} toolbox as the simulation platform. To solve the Master equation in \eqref{mastereq}, three quantities are needed: Hamiltonian (H), Lindblad operator ($L_j$) and the initial state of the qubit(s) ($\rho$), as inputs. In this part, we define  

\textbf{Hamiltonian:} For system Hamiltonian, we make use of the approximation in \eqref{Happrox}. As Hamiltonians are the description of the coherent part of the evolution, we take $H$ to be equal to the unitary matrices representing different gates. For example, to simulate the Hadamard gate, we take 
\begin{equation*}
H = (1/\sqrt{2})\begin{bmatrix}1 & 1 \\1 & -1\end{bmatrix}
\vspace{-2mm}
\end{equation*}

\textbf{Lindblad operators:} Lindblad operators are critical for decoherence analysis as they define the effect of environment on the system. As noted by Nielsen and Chuang \cite{chuang}, amplitude damping and phase damping are ideal models of noise that capture most of the important features of the noise occurring in quantum mechanical systems. Therefore, we confine our analysis to amplitude damping and phase damping in this paper. 

\textbf{Amplitude damping:} Amplitude damping is the class of quantum noise that accounts for the effect of loss of energy from the quantum system. The different physical realization of qubits have a different mechanism for energy dissipation e.g., in trapped-ion based quantum computer energy dissipation may occur as the emission of a photon. However, amplitude damping works as a general description of these different energy dissipation processes. 

To capture the effect of amplitude damping, a suitable Lindblad operator has to be defined. This energy dissipation process is described by the Lindblad operator for a single qubit
\begin{equation}
L_{amp} = \sqrt{\gamma}\hat{a} = \sqrt{\gamma}\begin{bmatrix}0 & 1 \\0 & 0 \end{bmatrix}
\vspace{-2mm}
\end{equation}
where $\gamma$ is the rate of spontaneous emission. The value of $\gamma$ defines the strength of the environment's effect on the system. $\hat{a}$ is the annihilation operator or lowering operator.

\begin{figure}[tb]
\centerline{\includegraphics[width=2.5in]{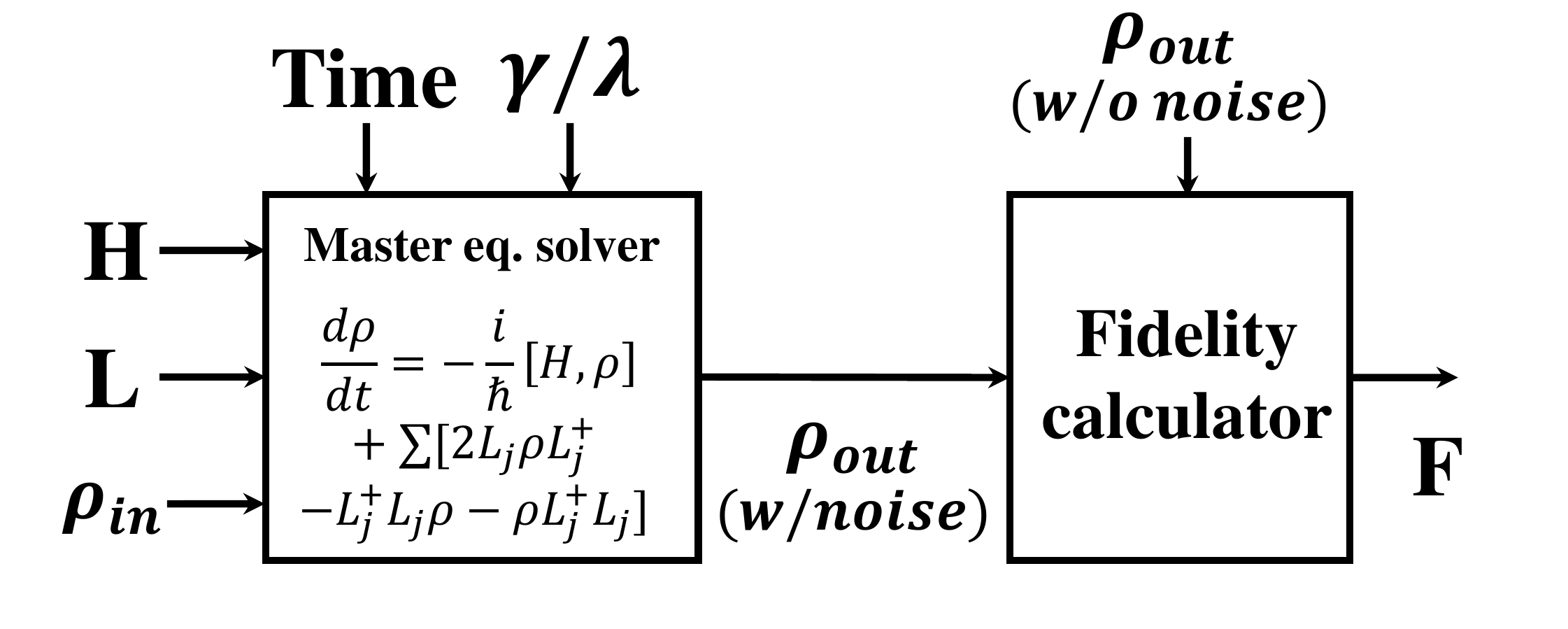}}
\caption{Diagram for the simulation framework.}
\label{box}
\vspace{-5mm}
\end{figure}

For a multi-qubit system, the Lindblad operator has to be extended from its single qubit form. For example, a three-qubit system will have following Lindblad operators where I is $2 \times 2$ identity matrix and $\otimes$ denotes tensor product: 
\begin{align}\label{ampdamp1}
L_{amp}^{(1)} = L_{amp} \otimes I \otimes I\\
L_{amp}^{(2)} = I \otimes L_{amp} \otimes I\\\label{ampdamp3}
L_{amp}^{(3)} = I \otimes I \otimes L_{amp}
\end{align}

\textbf{Phase damping:} Phase damping is the loss of quantum information without the loss of energy. With phase damping, qubit does not make a transition in the $\ket{0}$ and $\ket{1}$ states. The Lindblad operator for phase damping is given by the following equation where $\sigma_z$ is the \textit{Pauli-Z} operator and $\lambda$ is the parameter defining the strength of phase damping: 
\begin{equation}
L_{phase} = \sqrt{\lambda}\sigma_z = \sqrt{\lambda}\begin{bmatrix}1 & 0 \\0 & -1 \end{bmatrix}
\vspace{-2mm}
\end{equation}
The single qubit $L_{phase}$ operator can be expanded for multi-qubit case following the same method as in \eqref{ampdamp1}-\eqref{ampdamp3}.

\textbf{Input state:} Different input states, from pure states to mixed states, are simulated. Input states are fed into the Master equation as density matrix. For multi-qubit case, a state vector is first calculated for the composite system using tensor product among each qubit state vector ($\ket{\psi_{system}}=\ket{\psi_{qubit-1}} \otimes \ket{\psi_{qubit-2}} ... \otimes \ket{\psi_{qubit-n}}$). Then, the \eqref{dm} is used to prepare the input density matrix. The probability $p_i$ in \eqref{dm} is accounted for in the probability amplitude of the basis vectors $\ket{0}$ and $\ket{1}$ of the individual state vector (a and b in $\psi$). 
\vspace{-2mm}

\section{Single Quantum Gates}
In this section, we present the simulation results and note the insights. In \ref{sqg}, we discuss the noise effect of single quantum gates and in \ref{multi} we extend the analysis to several quantum circuits. 

We consider six quantum gates, CNOT, Hadamard, Toffoli, Fredkin, phase, and T-gate, in our analysis. Each quantum gate is represented by a unitary matrix \cite{branco} $U$ such that $UU^{\dagger} = I$ (conversely, any unitary matrix can be a quantum gate). The unitary matrix descriptions can be found in \cite{chuang}. 
CNOT is the control-NOT gate that flips the state of target bit if control-bit is 1. Toffoli is control-control-NOT; that means target bit will flip if both control bits are 1. 
Fredkin gate is control-SWAP. 
Hadamard, phase and T gates, however, are quantum-mechanical in nature and do not have classical analogue. Hadamard gate transforms the basis states to superposition states. Phase and T gate add phase to the state of a qubit.  

\subsection{Simulation setup}\label{sqg}
\begin{figure}[b]
\vspace{-5mm}
\centerline{\includegraphics[width=2.3in]{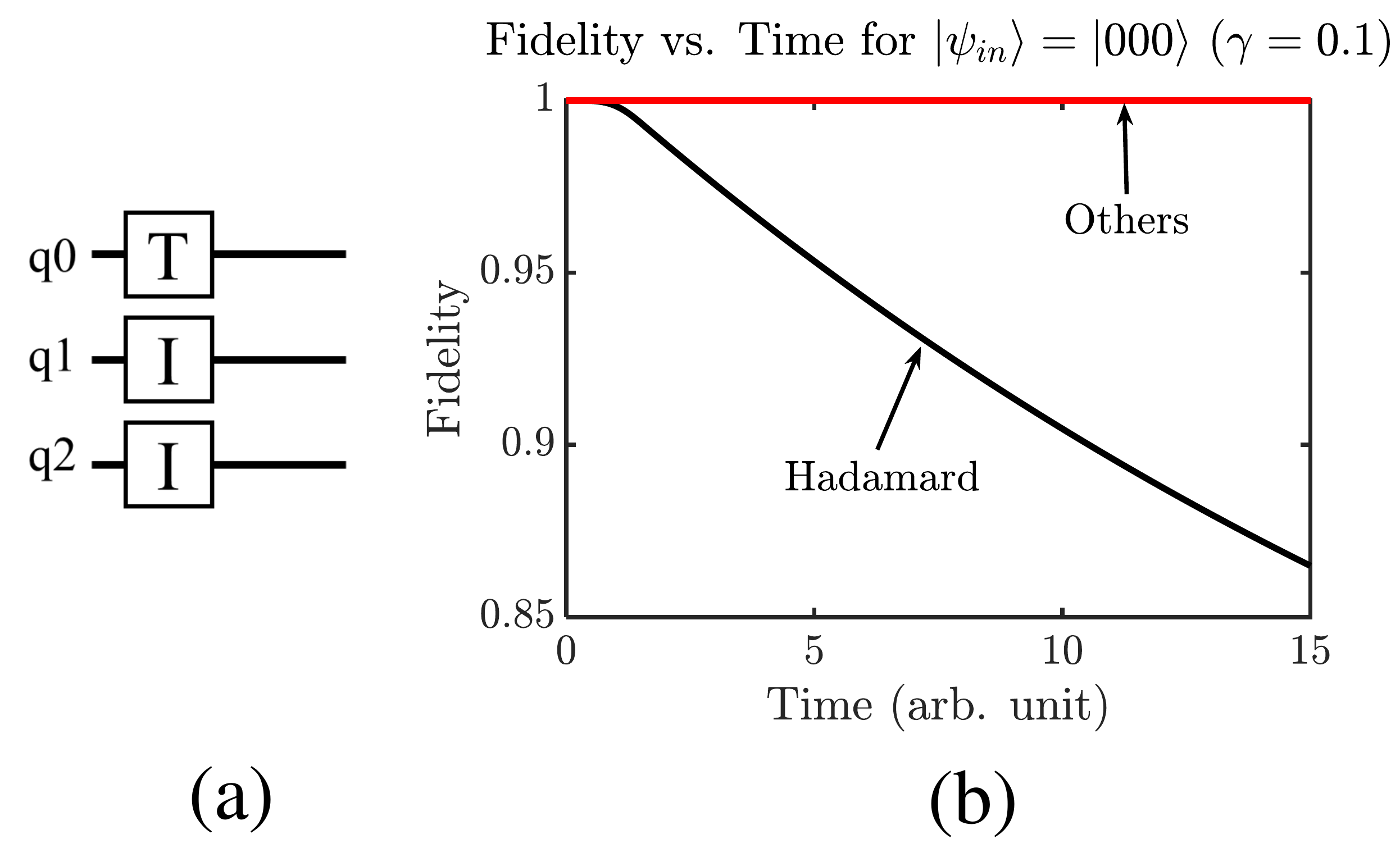}}
\caption{(a) Extending single-qubit operation to multi-qubit and (b) effect of amplitude damping on different gate operations for input = $\ket{000}$.}
\label{amp000}
\end{figure}

We systematically simulate the six different quantum gates with amplitude and phase damping effects. These gates work on a different number of qubits e.g. Hadamard gate works on single qubit while Toffoli and Fredkin are three-qubit gates. To make the comparison even, each gate has been extended to a common three-qubit system. For one-qubit gates like Hadamard, phase and T-gate, the respective gates work on a single qubit (say, q0) and two Identity operators work on remaining two qubits (say, q1 and q2) (Fig. \ref{amp000}a). Therefore, the Hamiltonian will be (e.g., considering T-gate)
\begin{align}
H = T \otimes I \otimes I = \begin{bmatrix}1 & 0 \\0 & e^{i\pi/4}\end{bmatrix}\otimes \begin{bmatrix}1 & 0 \\0 & 1\end{bmatrix} \otimes \begin{bmatrix}1 & 0 \\0 & 1\end{bmatrix}
\end{align}
Similarly, for two-qubit gates, Identity operator will work on the remaining qubit (e.g. for one possible case where CNOT gate is working on q1 and q2, the Hamiltonian will be $H = I \otimes CNOT$). 
For three-qubit gates in this three-qubit system, no such extension is necessary. However, for more than three-qubit system, the three-qubit gates can be extended with the aid of Identity operator as necessary. 


The effect of decoherence is shown through the fidelity vs. time graphs.  It is to be noted that time is in arbitrary unit scale. Different physical representation of quantum computer has different time-scale. A list can be found in \cite{chuang}. For a particular physical device, the time-axis may just be considered with a constant multiplier. As mentioned in \ref{basics}, fidelity measures the distance between two states. To calculate the fidelity between following two states: one of the states is the density matrix with noise (say, $\rho_{out}$) i.e. the actual output and the other is the density matrix without noise (say, $\rho_{target}$) i.e. the state in the ideal case. $\rho_{target}$ is calculated using the Schr\"{o}dinger equation without any noise. 

\subsection{Results with amplitude damping} 
First, we discuss the amplitude damping trend. It is to be noted, a general characteristic of quantum operation is that a set of states may remain unchanged under the operation \cite{chuang}. The Lindblad operator for amplitude damping contains an annihilation operator which will \textit{lower} the state $\ket{1}$ to state $\ket{0}$ keeping $\ket{0}$ unchanged. Therefore, it can be predicted that if there is $\ket{1}$ in the output state, it will experience damping and fidelity will be low. The more the $\ket{1}$ components in the density matrix at any time instance, the more will be the damping. With no $\ket{1}$s in the density matrix (i.e. all $\ket{0}$s), the output with noise should be equal to the ideal case i.e. output without noise. The simulation results corroborate to this intuition. 

\begin{figure*}[t]
\centering
\subfloat[]{\includegraphics[width=1.7in]{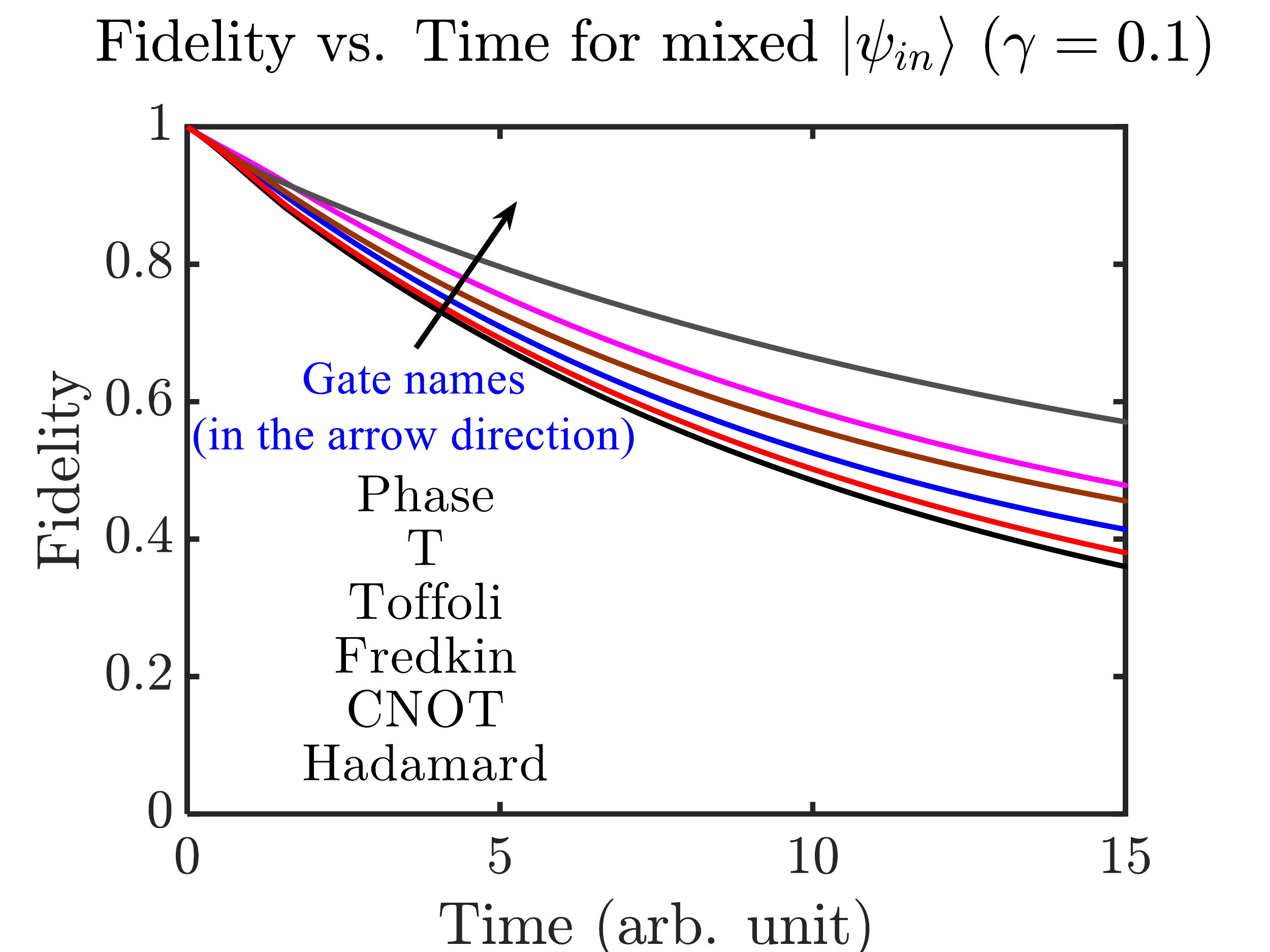}
\label{mixedamp}}
\hfil
\subfloat[]{\includegraphics[width=1.7in]{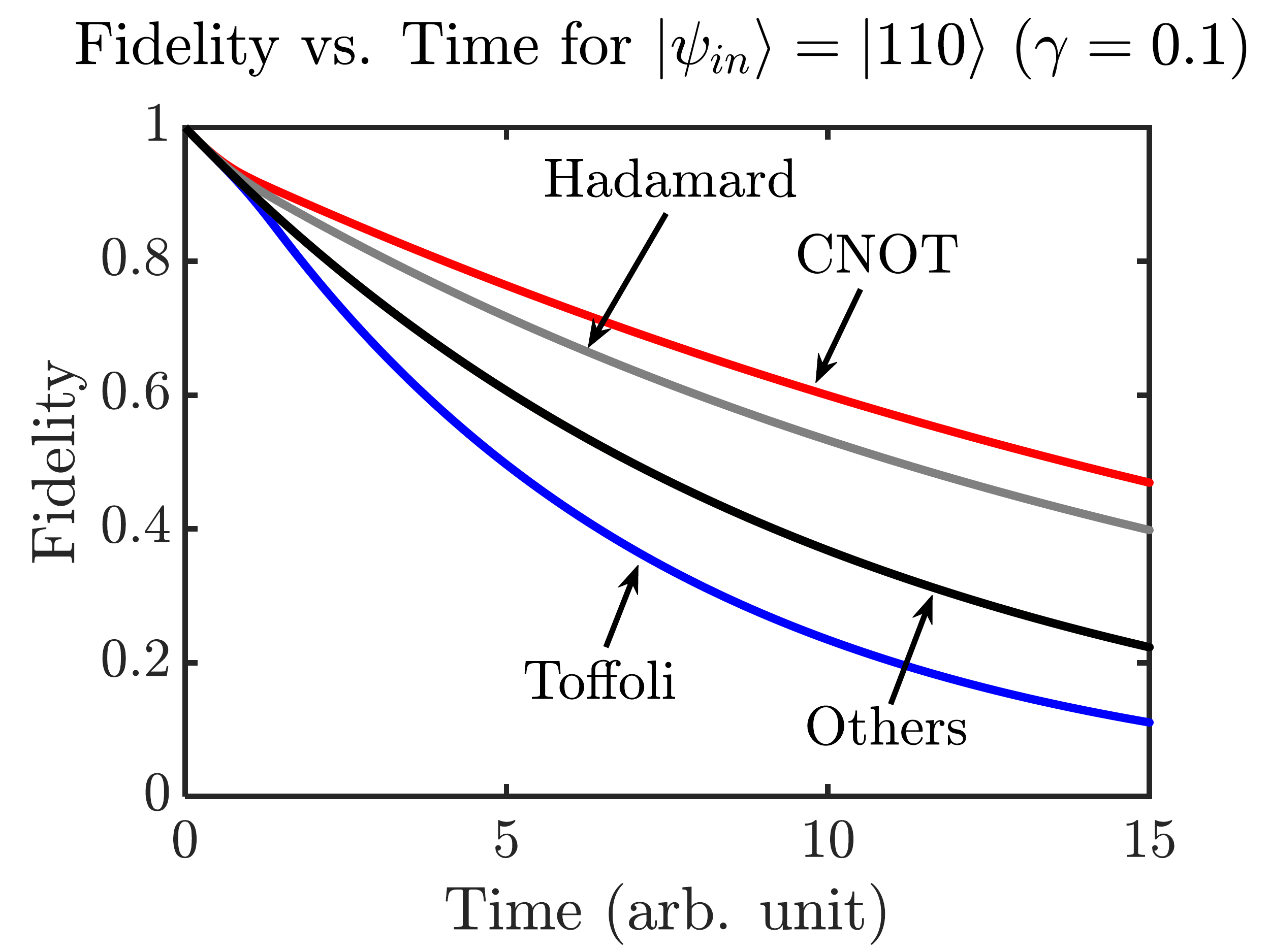}
\label{amp110}}
\hfil
\subfloat[]{\includegraphics[width=1.7in]{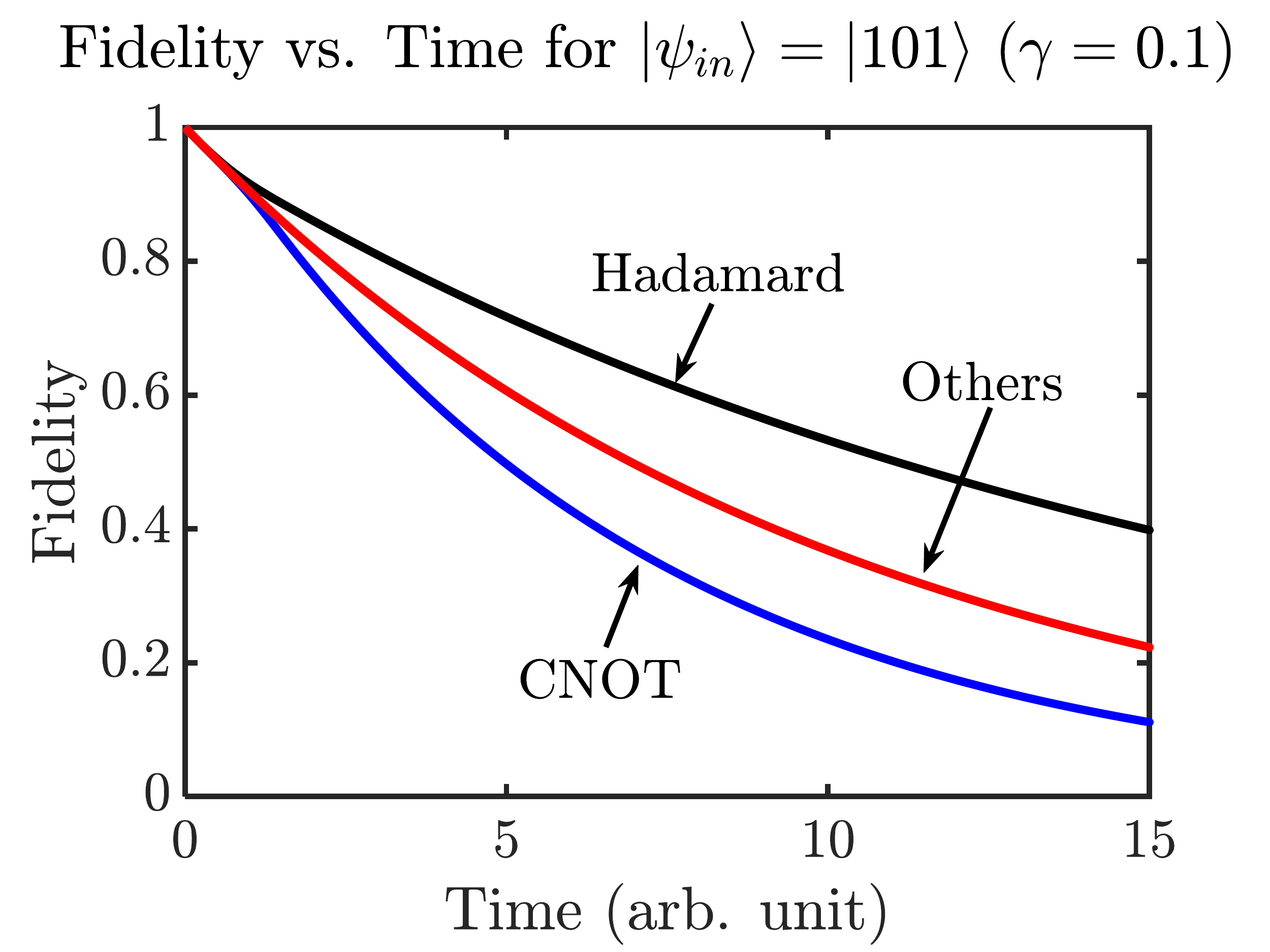}
\label{amp101}}
\hfil
\subfloat[]{\includegraphics[width=1.7in]{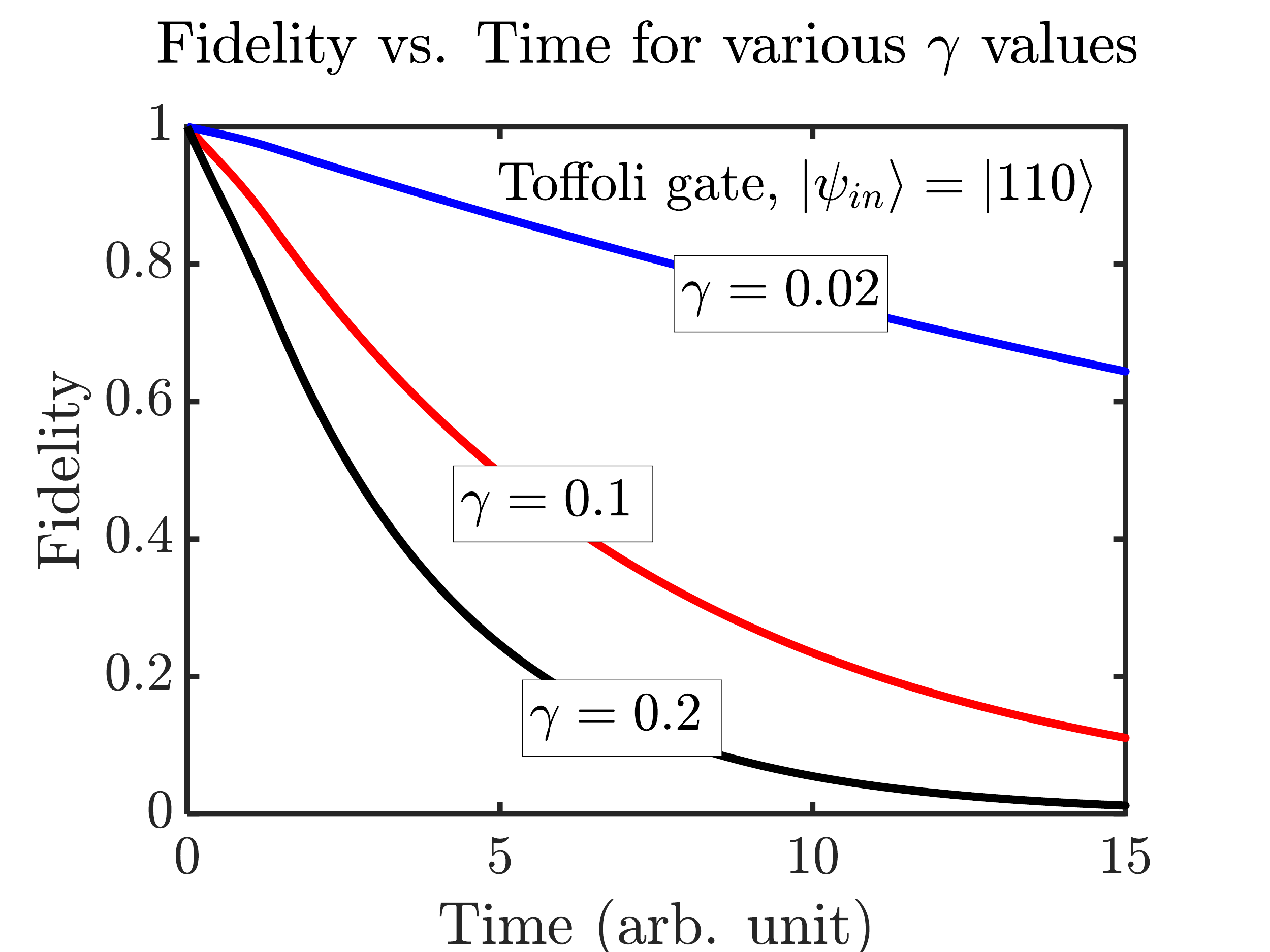}
\label{ampdiffgamma}}
\caption{Amplitude damping simulation. (a) input is a superposition state, (b) input = $\ket{110}$. Toffoli shows the highest decoherence, (c) input = $\ket{101}$. CNOT shows the highest decoherence, and (d) fidelity for different values of $\gamma$. A higher values results in more loss in fidelity.}
\label{fig_sim}
\vspace{-5mm}
\end{figure*}

\begin{figure*}[tb]
\centering
\subfloat[]{\includegraphics[width=1.7in]{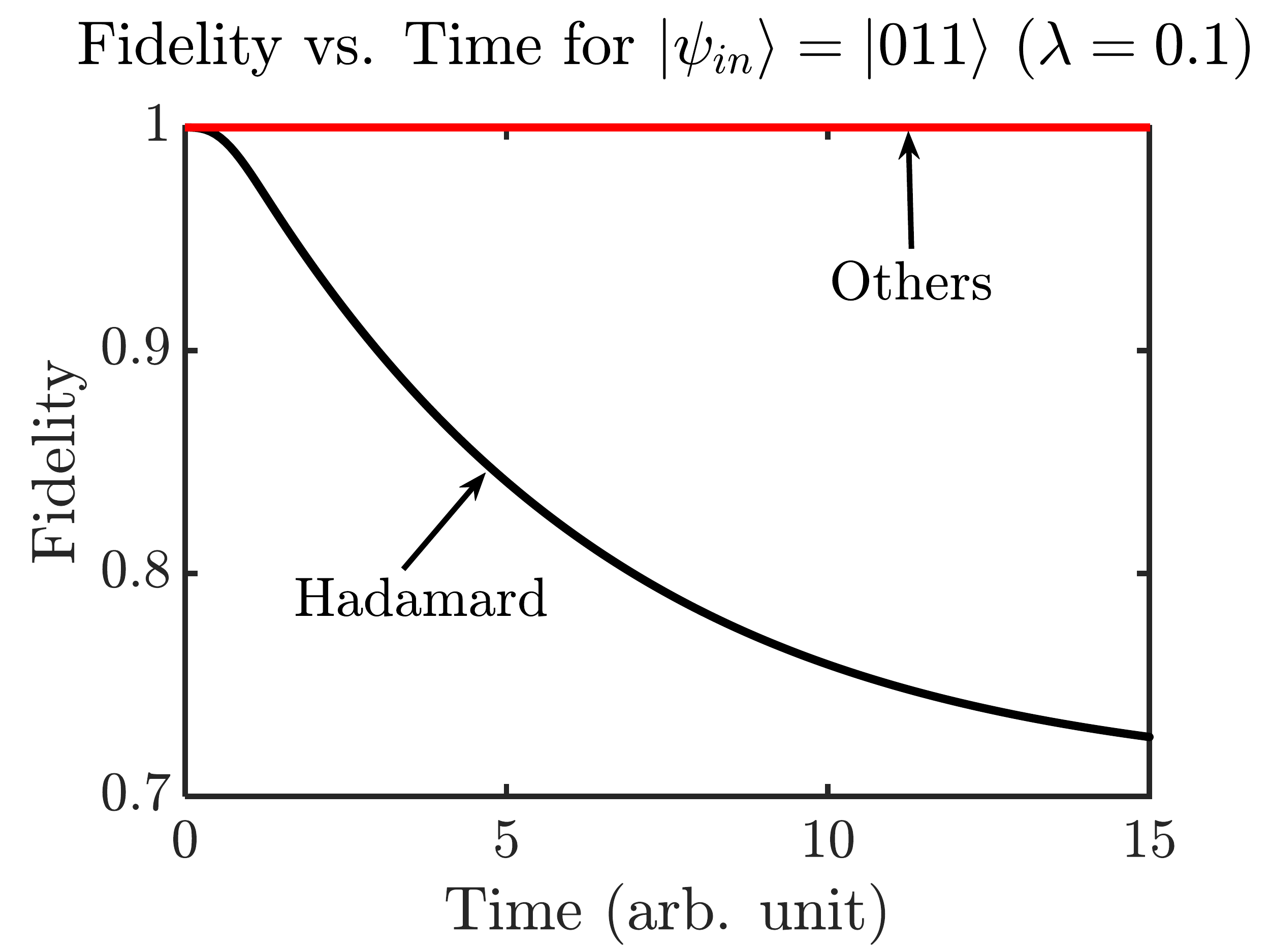}
\label{phase011}}
\hfil
\subfloat[]{\includegraphics[width=1.7in]{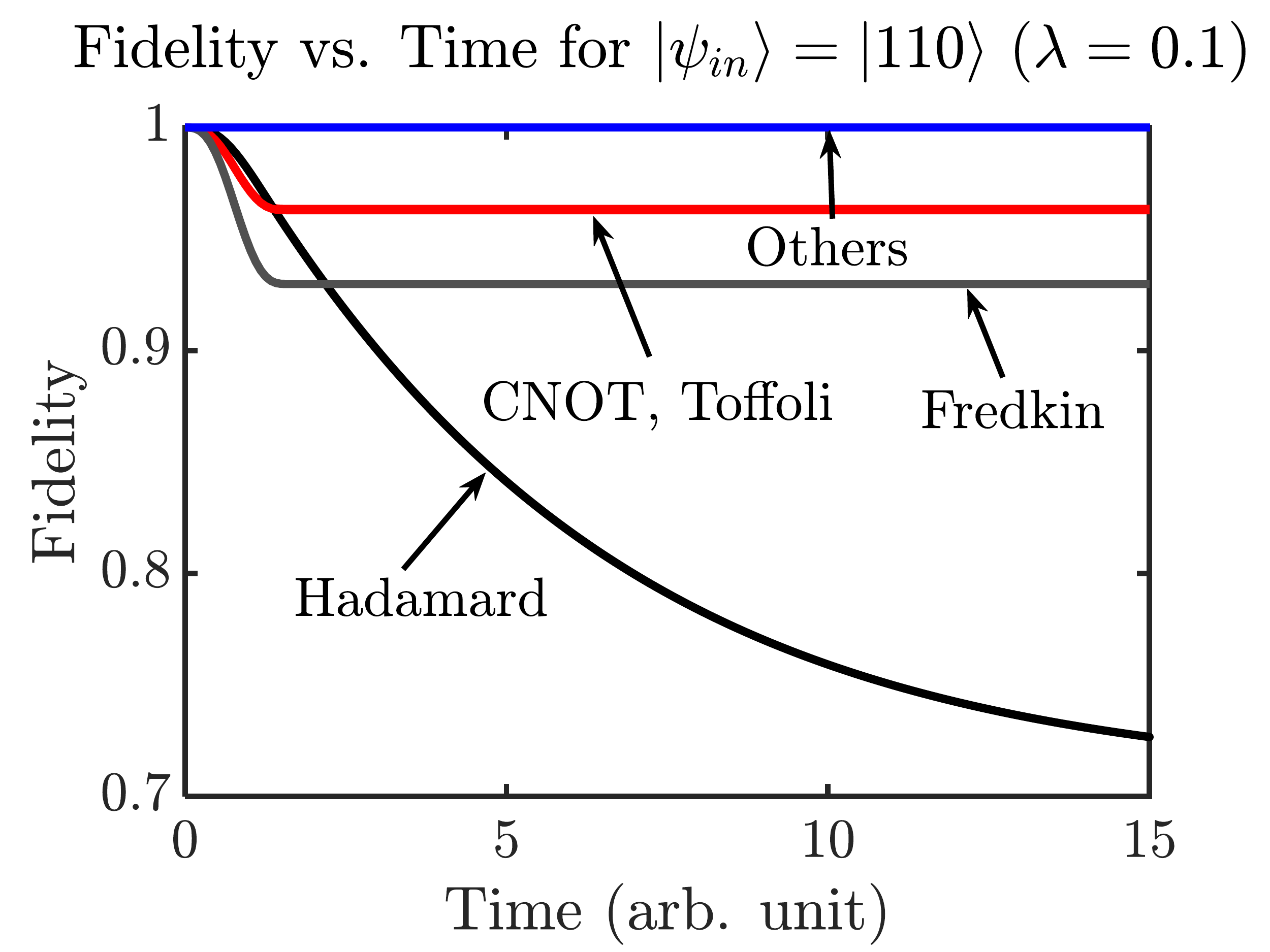}
\label{phase110}}
\hfil
\subfloat[]{\includegraphics[width=1.7in]{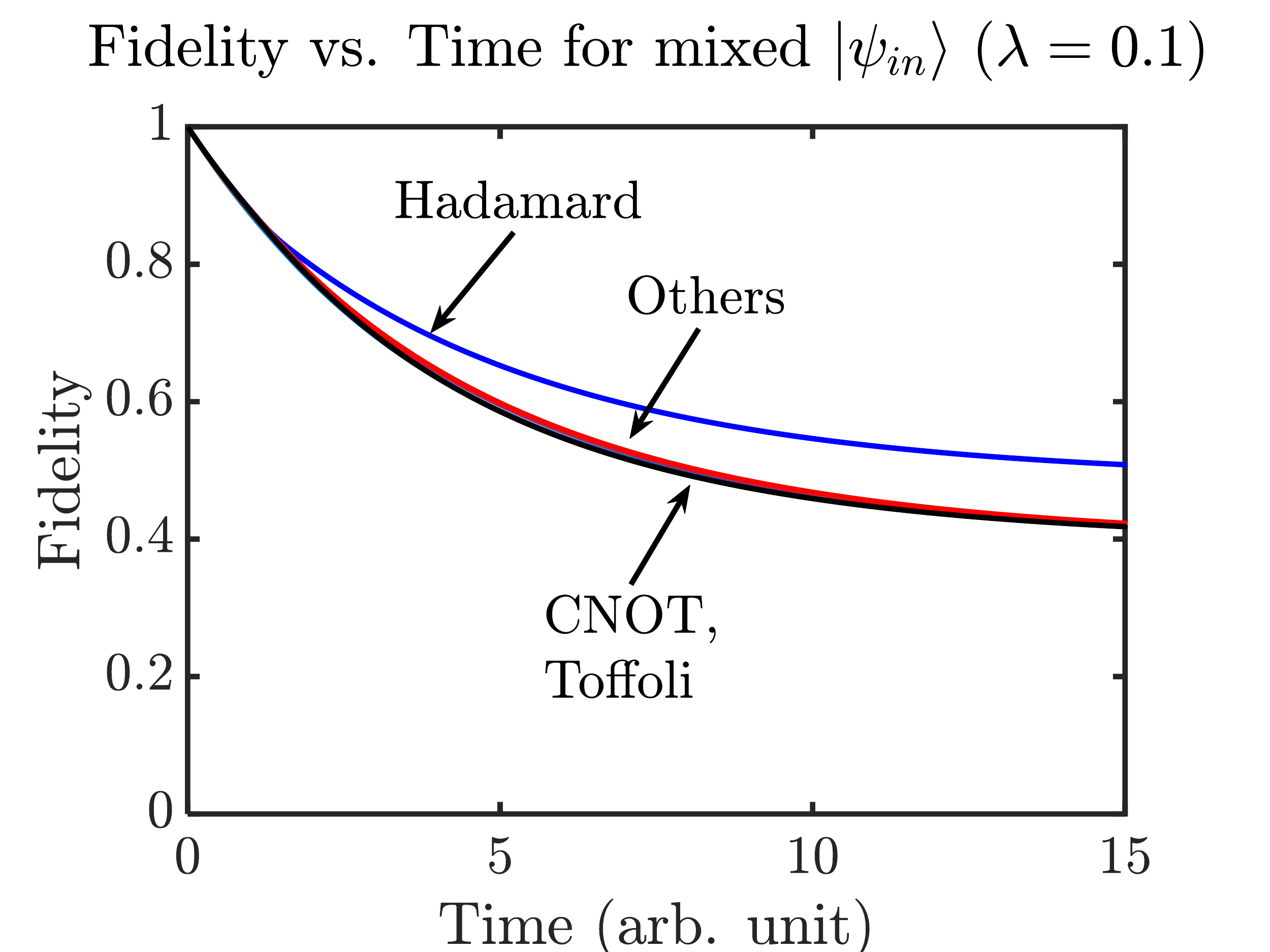}
\label{pahsemixed}}
\hfil
\subfloat[]{\includegraphics[width=1.7in]{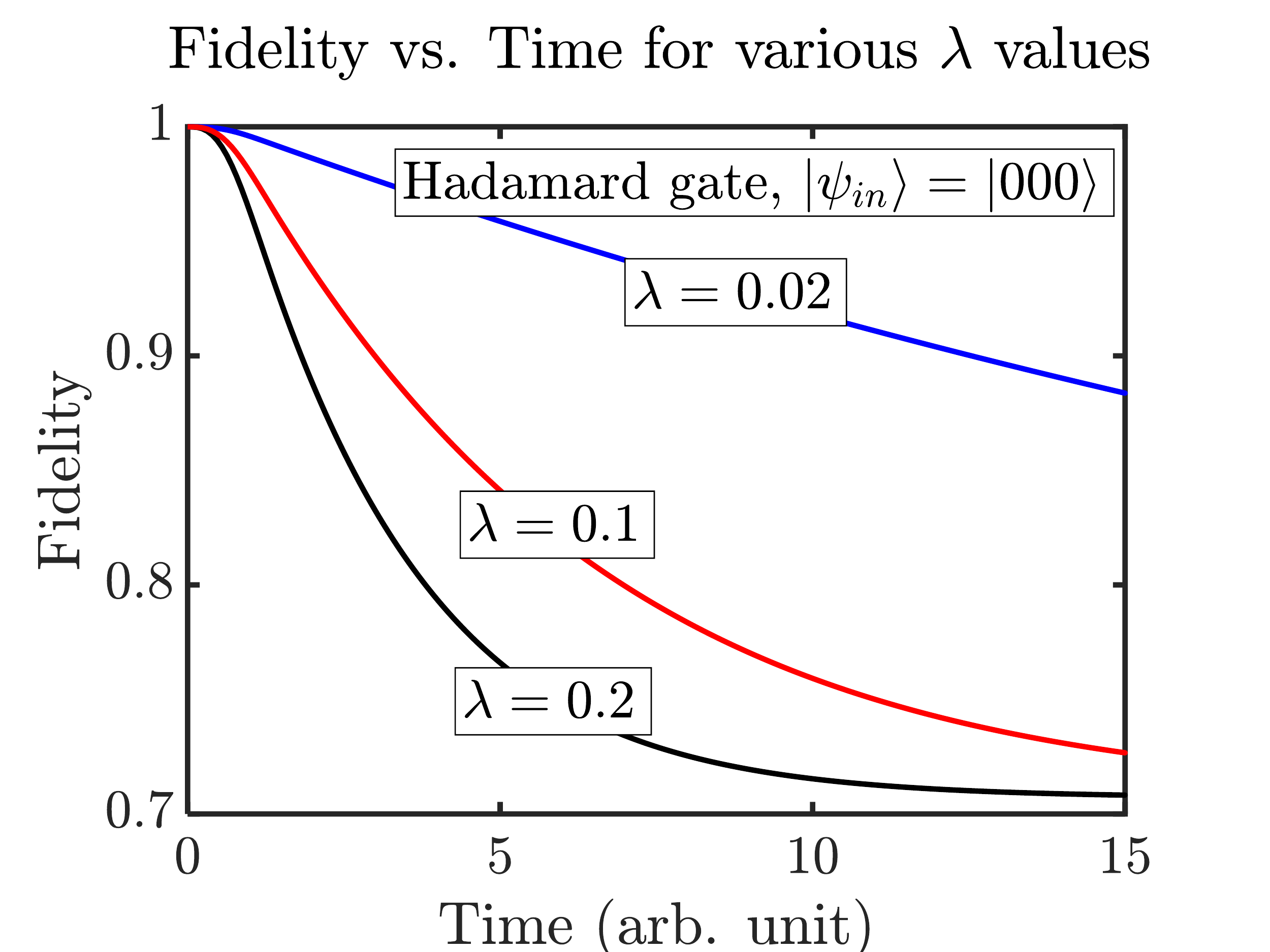}
\label{phasedifflambda}}
\caption{Phase damping results. (a) input = $\ket{011}$, (b) input = $\ket{110}$; CNOT and Fredkin operation loses fidelity as the input qubit states change due to gate operation, (c) input is a mixed state; all the gates show decaying trend and, (d) effect of the value of $\lambda$ on the fidelity; higher $\lambda$ causes more decoherence.}
\label{phase}
\vspace{-5mm}
\end{figure*}
Fig. \ref{amp000}b shows the fidelity vs. time plot for $\ket{\psi} = \ket{000}$ case. Except Hadamard all other gates under consideration exhibit ideal fidelity = 1. The reason behind this needs to be understood. For an input of $\ket{000}$, the gate operations under consideration (CNOT, Toffoli, Fredkin, phase and T-gate) results in an output of $\ket{000}$. As there are no $\ket{1}$s at any time instance of the state, amplitude damping leaves the state invariant of noise. Hadamard gate is an exception to this since it transforms a pure $\ket{0}$ or $\ket{1}$ state to the superposition state of both $\ket{0}$ and $\ket{1}$. 
\begin{align*}
\ket{0} \xrightarrow{Hadamard} (\ket{0} + \ket{1})/\sqrt{2}\\
\ket{1} \xrightarrow{Hadamard} (\ket{0} - \ket{1})/\sqrt{2}
\end{align*}
Therefore, Hadamard operation introduces a $\ket{1}$ with $50\%$ probability in the state which is susceptible to amplitude damping and thus, the fidelity starts to degrade. In other words, the qubit state starts to decohere. Similar effects can be observed if the input is in the mixed state instead of pure basis states. Fig. \ref{mixedamp} shows the matching trend. When the input state is mixed i.e. superposition of both $\ket{0}$ and $\ket{1}$, all the gates under consideration show decoherence due to amplitude damping. 

It is helpful to investigate a few other cases to completely understand the trend. Consider, $\ket{\psi_{in}} = \ket{110}$. For this input, the output of the Toffoli gate shows the highest decoherence and CNOT shows the lowest (Fig. \ref{amp110}). Toffoli is a controlled-controlled-NOT operation which means if the first two qubits are $\ket{1}$ then the third qubit will flip the state. Therefore, after Toffoli operation, the output is $\ket{111}$. Due to the maximum number of $\ket{1}$s, the output of the Toffoli gate suffers maximum decoherence among other gate outputs. CNOT flips the target qubit (q1 in this example) if the control qubit (q0) is $\ket{1}$. After the CNOT operation, the output will be $\ket{100}$ with a single $\ket{1}$. Therefore, the decoherence is lowest among other gate outputs. Again, for $\ket{\psi_{in}} = \ket{101}$, CNOT outputs a $\ket{111}$. Therefore, its output has the highest decoherence (Fig. \ref{amp101}) and it is similar to the Toffoli gate with $\ket{\psi_{in}} = \ket{110}$. From this discussion, one insight is that the decoherence primarily depends on the states of the qubit(s). If two different gate operations evolve a set of qubits to the same states, then the decoherence behavior will be same (given that the environment and qubit interaction does not face temporal variation during two gate operations). 

Finally, the impact of the value of $\gamma$ is shown in Fig. \ref{ampdiffgamma}. As an example, we simulate Toffoli gate for $\ket{\psi_{in}} = \ket{110}$ for three different $\gamma$ values ($\gamma = 0.02, 0.1, 0.2$). $\gamma = 0.02$ and $\gamma = 0.2$ are dubbed as weak and strong amplitude damping respectively\cite{specialissue}. Additionally, we simulated for $\gamma = 0.1$ as a middle-ground value. Evidently, a higher $\gamma$ value results in higher decoherence within the given time. It is to be noted that the value of $\gamma$ must be experimentally calibrated for a particular physical quantum computer.

\subsection{Results with phase damping} 
As phase damping is a completely quantum mechanical phenomenon, 
sometimes it is necessary to resort to pure mathematical treatment. However, from the simulation results presented in Fig. \ref{phase}, valuable insights can be garnered. 

Mathematically, phase damping affects the off-diagonal elements of a density matrix and decays those to zero with time. Therefore, if the density matrix representation of the qubits states does not have any off-diagonal during the evolution then phase damping will not affect. There are two possible cases when off-diagonal terms may appear in the density matrix: (i) if the input of a quantum gate is a superposition state and/or (ii) qubit states undergo change due to gate operation. 

The preceding statements can be validated with the simulation results. First, consider, the qubits start from a pure state $\ket{011}$. It can be easily verified, that none of the gate operations (except Hadamard which is discussed later) will change this state. For example, CNOT (considering, q0 as the control bit and q1 as the target bit) will not flip the target bit as control bit is $\ket{0}$. Same is true for Toffoli (control-control-NOT) and Fredkin (control-SWAP). Phase and T gates do not alter basis states, hence $\ket{011}$ will remain unchanged after the gate operation. As the initial state is not undergoing change due to the gate operations, no off-diagonal terms will appear at any instance and hence the state will not be affected due to phase damping. Therefore, the fidelity will remain $1$ throughout (Fig. \ref{phase011}). However, Hadamard gate (working on q0) does transform the pure states to superposition state. As the state changes, off-diagonal terms appear at the density matrix due to Hadamard gate operation and the qubits state experience decoherence due to phase damping. Therefore, the fidelity of Hadamard gate operation shows a decaying trend. Hadamard gate operation on $\ket{0}$ introduces off-diagonal terms, this can be checked using the following mathematical routine where $U_H$ is unitary transformation associated with Hadamard gate and $\rho_{in}$ is the density matrix representation of basis vector $\ket{0}$:
\begin{align*}
\rho_{out} &= U_{H} \rho_{in} U_{H}^{\dagger} \\
&= \frac{1}{\sqrt{2}} \begin{bmatrix}1 & 1 \\1 & -1 \end{bmatrix} \begin{bmatrix}1 & 0 \\0 & 0 \end{bmatrix} \frac{1}{\sqrt{2}} \begin{bmatrix}1 & 1 \\1 & -1 \end{bmatrix}
= \begin{bmatrix}0.5 & 0.5 \\0.5 & 0.5 \end{bmatrix}
\end{align*}


Next consider, $\ket{\psi_{in}} =\ket{110}$. As the control bit(s) is(are) now $\ket{1}$($\ket{11}$), CNOT, Toffoli and Fredkin gate will change the qubit states to $\ket{100}$, $\ket{111}$ and $\ket{101}$ respectively. Therefore, during this evolution, off-diagonal terms will appear in the density matrix and phase damping will kick-in. It is evident from Fig. \ref{amp110} as fidelities after CNOT, Toffoli, and Fredkin gate operation are less than 1. However, the fidelity becomes constant after a certain time (t = $\pi/2$ unit). This happens because at and after that time, the off-diagonal terms in the density matrix vanishes and therefore, phase damping ceases to have any impact. 
Hadamard gate operation shows similar trend as in previous case for same reason. 


Next mixed state case is discussed. If the qubits start from a mixed state which is neither pure basis state vectors $\ket{0}$ nor $\ket{1}$, rather a superposition of both (e.g. $\ket{\psi}=0.8\ket{0} + 0.6\ket{1})$ then the density matrix will have off-diagonal terms ($\rho = \ket{\psi} \bra{\psi} = \begin{bmatrix}0.64 & 0.48 \\0.48 & 0.36 \end{bmatrix}$). Therefore, qubits will undergo decoherence due to phase damping. Fig. \ref{pahsemixed} shows that fidelity is decaying for every gate operation. 

Finally, Fig. \ref{phasedifflambda} shows the effect of varying $\lambda$ value which defines the strength of phase damping on fidelity. As expected, a higher $\lambda$ value results in sharper decay. Although the effect is shown only for Hadamard gate for $\ket{000}$ due to space limit, the trend will be similar for other gates and other input states. 
\section{Multi-depth Quantum Circuits}\label{multi}
\begin{figure}[tb]
\centerline{\includegraphics[width=3in]{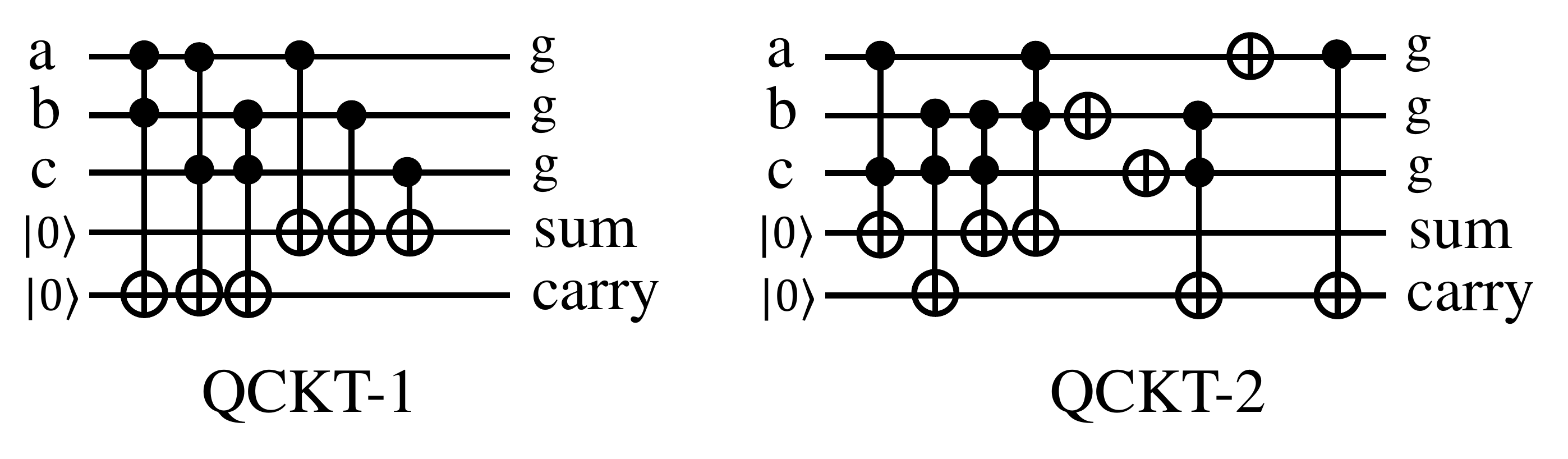}}
\caption{Quantum full adder circuits from RevLib \cite{revlib}. a and b are the two input bits and c is the carry-in. $g$ stands for garbage output.}
\label{adder}
\vspace{-3mm}
\end{figure}

Two different implementations of reversible full adders (Fig. \ref{adder}) are simulated. The adder quantum circuits are based on NOT, CNOT and Toffoli gates. Both have two ancilla qubits (ancilla qubit is auxiliary qubit that is independent of values of inputs and is necessary to execute a function such as addition in this case). However, QCKT-1 has 6 sequential gate operations whereas QCKT-2 has 9 sequential gate operations.  Therefore, the depth of circuits can be taken to be 6 and 9 respectively as they require 6 and 9 time-steps to completely evolve the input qubit state to the target output state.

\begin{figure}[tb]
\centerline{\includegraphics[width=1.8in]{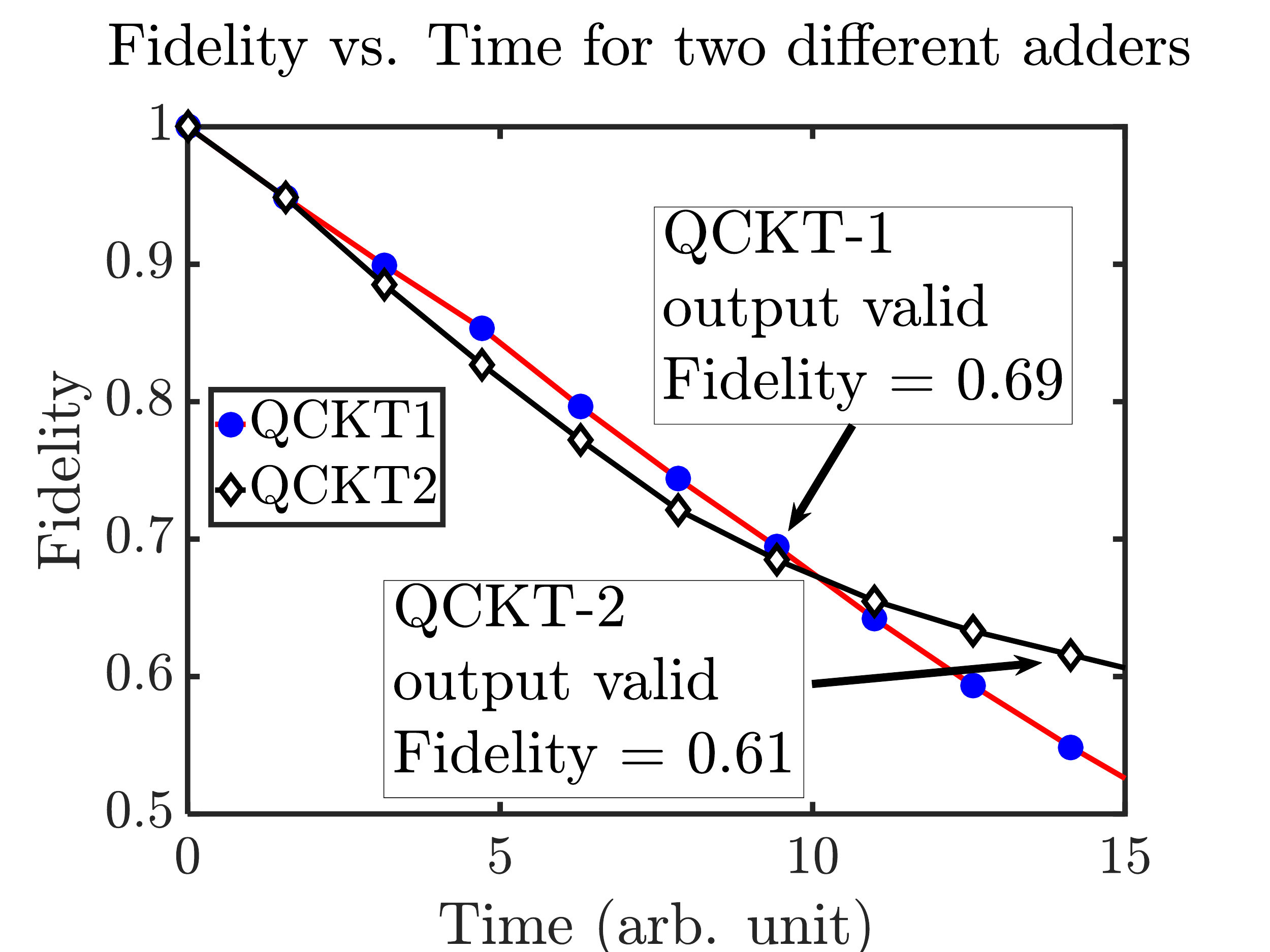}}
\caption{Adder fidelities. Lower-depth QCKT-1 exhibits better fidelity than higher-depth QCKT-2.}
\label{qcktadder}
\vspace{-5mm}
\end{figure}

Fidelity vs. Time for this multi-qubit and multi-depth quantum circuits are shown in Fig. \ref{qcktadder} for amplitude damping with $\gamma = 0.02$. As a test-case, initial qubit states are taken to be $\ket{abc} = \ket{111}$. In case of QCKT-1 six sequential gate operations finishes at sixth time-step. Therefore, the qubit states can be measured (read-out) at that point. For QCKT-2, the read-out can be done at or after nine time-steps. QCKT-1 shows better fidelity than QCKT-2. As QCKT-2 takes longer than QCKT-1 to finish gate operations, qubits decohere longer in QCKT-2 than QCKT-1. \textit{In general, the higher the depth of the quantum circuit, the greater the decoherence.} 

From Fig. \ref{qcktadder}, it might be tempting to conclude that after a certain time-step (about 7, for this case) the fidelity behavior of QCKT-2 gets better than its lower-depth counterpart QCKT-1. However, this arises due to the input state of this test-case. Note that from \ref{sqg}, amplitude damping affects the qubit states with more $\ket{1}$s. For QCKT-1 for an input of $\ket{111}$ ($\ket{11100}$ with ancilla qubits), the ideal output is $\ket{11111}$ whereas for QCKT-2 the output is $\ket{00011}$. More number of $\ket{1}$s in case of QCKT-1 causes it to decohere more with more time. However, over the complete range of, at least, eight classical input combinations, the average fidelities of lower-depth quantum circuits are better than higher-depth ones. For the adder circuits of the discussion average fidelities are 0.74 and 0.59 for QCKT-1 and QCKT-2 respectively (lower-depth QCKT-1 has $20\%$ better fidelity than higher-depth QCKT-2).


The takeaway from the preceding discussion is that while designing quantum circuits designer should focus to reduce the depth of the circuit to achieve better fidelity. 

\section{Conclusion}
We performed fidelity analysis of multi-qubit and multi-depth quantum circuits using amplitude and phase damping as decoherence mechanisms. Our study indicates that decoherence is primarily data dependent and simple optimization of logic with lower-depth circuits can improve the fidelity of the system by 20\%. 

\section*{Acknowledgment}
This work is supported by SRC (2847.001), NSF (CNS- 1722557, CCF-1718474, DGE-1723687 and DGE-1821766) and DARPA Young Faculty Award (D15AP00089).

\end{document}